# The peculiar shapes of Saturn's small inner moons as evidence of mergers of similar-sized moonlets

A. Leleu[1,2], M. Jutzi[1], M. Rubin[1]

1. Physics Institute, Space Research and Planetary Sciences, Center for Space and Habitability - NCCR PlanetS - University of Bern, Switzerland
2. IMCCE, Observatoire de Paris - PSL Research University, UPMC Univ. Paris 06, Univ. Lille 1, CNRS, 77 Avenue Denfert-Rochereau, 75014 Paris, France
Email : adrien.leleu@space.unibe.ch

**The Cassini spacecraft revealed the spectacular, highly irregular shapes of the small inner moons of Saturn[1], ranging from the unique "ravioli-like" forms of Pan and Atlas[2,3] to the highly elongated structure of Prometheus. Closest to Saturn, these bodies provide important clues regarding the formation process of small moons in close orbits around their host planet[4], but their range of irregular shapes has not been explained yet.**

**Here we show that the spectrum of shapes among Saturn's small moons is a natural outcome of merging collisions among similar-sized moonlets possessing physical properties and orbits that are consistent with those of the current moons. A significant fraction of such merging collisions take place either at the first encounter or after 1-2 hit-and-run events, with impact velocities in the range of 1-5 times the mutual escape velocity. Close to head-on mergers result in flattened objects with large equatorial ridges, as observed on Atlas and Pan. With slightly more oblique impact angles, collisions lead to elongated, Prometheus-like shapes. These results suggest that the current forms of the small moons provide direct evidence of the processes at the final stages of their formation, involving pairwise encounters of moonlets of comparable size[4,6,7].**

**Finally, we show that this mechanism may also explain the formation of Iapetus' equatorial ridge[8], as well as its oblate shape[9].**

The small inner moons Atlas, Prometheus, Pandora, Janus, and Epimetheus are repelled by the rings of Saturn at a rate that is proportional to their mass and decreases with their distance[11,4]. It has thus been proposed that these moons were formed in a pyramidal regime (i.e. by a series of mergers of similar sized bodies) as they migrated away from the rings[4,7]. This scenario is supported by the observations of the small inner Saturnian satellites, as bodies with similar semi-major axis have comparable masses: $m_{Prometheus}/m_{Pandora}$~1.16 and $m_{Janus}/m_{Epimetheus}$ ~3.6 (JPL SSD, not accounting for error bars), and the mass of these bodies increase with their distance to the rings. The pyramidal regime provides an alternative to the formation by gradual accretion of small aggregates of ring material onto a proto-moon. The later scenario would result in Roche ellipsoids[3,5], not consistent with the observed shapes of the small moons. For instance, while Atlas and Pan require a mechanism that makes their shapes flatter[5], Prometheus is over-elongated and its long axis extends beyond the Roche lobe[3].

Here we develop a model for the late stages of the formation of the small inner moons of Saturn from Pan to Janus and Epimetheus, assuming that they formed in the pyramidal regime[4,7]. We investigate if the shapes resulting from the final collisional mergers are consistent with the current shapes of these moons. To this end, we combine N-body simulations to estimate the possible range of impact angles and velocities between the precursors of a given moon, and Smooth Particle Hydrodynamics (SPH) simulations to obtain the outcome of the collisions.

For low-velocity impacts, which are expected given the range of eccentricities and semi-major axis of the small Saturnian satellites ($e<0.01$, $a\sim140000$ km), the shape resulting from a collision depends on the impact angle $\theta$, the angle between the relative position and relative velocity of the moonlets, and the ratio $V_{imp}/V_{esc}$ between the impact velocity and the mutual escape velocity[12]. Considering similar-sized moonlets of low density ($\rho = 700$ kg/m$^3$) and of semi-major axis $a = a_{Atlas}$, this ratio can be estimated as $V_{imp}/V_{esc} \sim 0.58 e/\mu^{1/3}$ (see Methods), where $e = max(e_1, e_2)$ is the eccentricity of the moonlets prior to collision, and $\mu = (m_1+m_2)/m_{Saturn}$ is the mass ratio between the colliding moonlets and Saturn. Using the current masses and eccentricities of the small moons, we obtain $V_{imp}/V_{esc} \sim 3.0$ for $m_1+m_2 = m_{Atlas}$, $\sim 1.9$ for $m_{Prometheus}$ and $\sim 3.9$ for $m_{Pandora}$. There is a priori no reason to believe that the eccentricity during the mergers was comparable to the current eccentricity of the moons. We show, however, that it was probably the case: although the collisional growth of the moonlets efficiently damped their eccentricity, and the oblateness of Saturn prevented their excitation through secular perturbation from the main moons (Methods), their eccentricity would increase prior to collision: as the semi-major axes of two moonlets get close, they enter a chaotic area due to the overlap of first-order mean-motion resonances[13,14]. We studied the evolution of the eccentricity in this area in the case of $\mu_{Atlas}$, $\mu_{Prometheus}$, and $\mu_{Janus}$, using numerical integration of the 3-body problem, taking into account the oblateness of Saturn (the effects of the rings and tides are negligible on the considered timescale[11,15,4])(see Methods). We found that the eccentricities rapidly reach $3\mu^{1/3} < e_{exc} < 9\mu^{1/3}$, values that are comparable to the current eccentricities of the small moons. As the excited eccentricity is proportional to $\mu^{1/3}$, this leads to collisions in the range $0 < V_{imp}/V_{esc} \leq 5$, regardless of the size of the considered moonlets. For Atlas-like objects, these calculations indicate typical impact velocities of order of a few 10 m/s.

We explore the outcome of such low-velocity (comparable to a few times the escape velocity) collisions among similar-sized moonlets using a three-dimensional SPH code[16,17,18] (Methods). Our code includes sophisticated models for material strength and porosity, effects not considered in previous studies[5,6,19]. We investigate a range of target-to-impactor mass ratios $m_1/m_2$, impact angles $\theta$, and velocities $V_{imp}$. The small moonlets are modeled as porous self-gravitating aggregates with densities of $\rho$ = 500 - 700 kg/m$^3$ and radii of $R_1 \sim 11$ km. Since the small moons are located very close to Saturn, at or within the Roche distance, tidal effects are significant[19] and are included (Methods). Our three-dimensional SPH calculations show that only collisions with a close to head-on impact angle lead to merging into a stable structure (Figures 1 and 2). Impacts with slightly larger impact angles lead to elongated, rotating structures which are split subsequently into two components. Because of the small Hill radius ($R_{Hill}/R_1 \sim 1.7$ at a distance of ~1.4 10$^8$ m assuming a density of ~ 700 kg/m$^3$), they quickly become unbound and separate. For even more oblique impact angles, collisions lead to hit-and-run events with not much interaction[20]. The transition from merging to non-merging collisions can be defined as a function of the angular momentum[12] as given by the relative velocity $V_{imp}$ and the impact angle $\theta$ (Figure 2).

For random coplanar orbits, the frequency function of the impact angle $\theta$ is given by $dP = \cos(\theta)\,d\theta$ (see Methods). This distribution is hence maximum for head-on collisions, which favors the merging of the moonlets. However, as the merging domain is rather narrow (Figure 2), it is probable that a moonlet pair undergoes a few hit-and-run collisions before the final merger, changing their mass ratio in the process. Directly prior to the first collision, and during the hit-and-run phase, the moonlets are on intersecting orbits in the inner part of the previously discussed chaotic area. Given that the time scale between collisions is significantly shorter than any perturbation (a few 10$^4$ years, see Methods), we assume that the gravitational interaction between the moonlets dominates their dynamics. To estimate their mass ratio at the time of merging, along with $V_{imp}/V_{esc}$ and $\theta$, we hence designed a model that takes into account the gravitational interactions between the two moonlets and Saturn, the oblateness of Saturn, and the collisions between the moonlets. When a collision occurs, $V_{imp}/V_{esc}$ and $\theta$ are compared to the domains defined in Figure 2 (a). In the case of non-merging collisions, the mass ratio $m_1/m_2$ after the collision is derived from the SPH collision model (Methods), and the new velocities are computed considering an elastic collision with dissipation[21] (Methods). We point out that in the case of Pan, we would need to model the rings as well to check if the eccentricity of the moonlets is excited to the same degree prior to collision. We note, however, that the chaotic area and the radial excursion corresponding to the considered eccentricities is comparable to the size of the Encke division (~300 km). Pan's precursors could hence have been on slightly eccentric orbits prior to collision without encountering ring material.

The model was run for different initial distributions of eccentricity, with $m_1/m_2=6/5$ and $m_1+m_2=m_{Atlas}$ as initial condition. However, it can be used to describe the formation of the other small inner moons as well, because for the size range of these small bodies, the results are scale-free for a given mean density and semi-major axis (Methods). In each case, we randomly selected $10^4$ initial conditions on intersecting orbits, such that 95% of the systems either merged or exceeded a threshold mass ratio within $10^5$ years. We set this threshold to $m_1/m_2\sim7/4$ as it is roughly the limit for which the formation of an equatorial ridge by merging of similar-sized bodies is still feasible (Figure 3). Because the size of the moonlet is comparable to its Hill sphere, ejections due to scattering events are extremely rare. For the systems that exceeded the threshold mass ratio before the final merger, we assume that the disruption/re-accretion processes occurring during the multiple collisions tend to produce featureless/random shapes within the Roche lobe, as it is the case for Pandora, Janus, and Epimetheus.

The distribution of $V_{imp}/V_{esc}$ and $\theta$ for the mergers occurring with $m_1/m_2<7/4$ are shown in Figure 2. These mergers occur typically after 1-2 merge-and-split or hit-and-run collisions. These collisions lead to a stable merged structure that result in a variety of shapes (Figure 3). Regardless of the chosen initial eccentricity distribution (Figure 2), close to head-on impacts lead to flattened structures with large-scale ridges, resembling the observed shapes of Atlas and Pan. Another frequent type of resulting structures is elongated shapes with characteristics similar to Prometheus, which result from impact parameters close to the boundary between merging and merge-and-split collisions. Depending on the chosen eccentricity distribution, 20 to 50% of the systems result in flat or elongated moons (these probabilities depend on the initial mass ratio between the precursors and increase as this mass ratio get closer to one). These results are consistent with the fraction of small moons possessing peculiar shapes: Pan, Atlas and Prometheus exhibit such features; while Pandora, Janus and Epimetheus do not. For completeness, the effect of inclination of the moonlets was also investigated. Inclination comparable to the current value for Atlas does not change significantly the ratio of successful merger, however it shifts $V_{imp}/V_{esc}$ toward lower values as it slightly increase the probability of hit-and-run encounters (see Methods).

The extended ridges observed on Saturn's small moons represent very smooth terrains, while the main bodies not covered by the ridge show more rocky/rough terrain[2]. In collisions, smooth terrain may result from ejecta reaccumulation as well as restructuring due to large deformations of the original surface (Methods). According to our model, for bodies resulting from ridge-forming mergers, such material (white in Figure 3) is mainly focused at the equatorial areas. The bulk densities resulting from the mergers are around $\rho$ = 400-600 kg/m$^3$ (Methods), in agreement with the observations[22]. These results suggest that the spectrum of structural features observed on Saturn's small moons – from flattened objects with extensive ridges to over-elongated shapes - can be explained by the final pairwise accretion of comparable-sized moonlets, supporting the pyramidal regime formation scenario[4,7], and suggest that no other significant process changed the shape of the small moons since the pyramidal regime. Our study also implies that hit-and-run collision between similar-sized objects were frequent throughout the formation of the small inner moons. Such an event was proposed to generate the F ring[6]. Although a hit-and-run event between Prometheus and Pandora was previously considered to explain the origin of the F ring, in our model the F ring material could originate from the hit-and-run collision of the precursors of Prometheus, then be trapped in the stability area between Pandora and Prometheus after the merging of the precursors.

The unique shapes of the small Saturnian moons are a consequence of the near-coplanarity of the system. The current inclinations of Jupiter's small moons, for example, do not favor head-on collision and hence make the merger of similar-sized moonlets inefficient. The similarity between the shapes of Amalthea and Prometheus, however, suggests a common origin for these objects, and would imply that the inclination of Amalthea's orbit evolved after its formation, perhaps crossing resonances with Io[23]. The mid-sized Saturnian moons have also been proposed to have formed in the pyramidal regime[4,7]. Although the features we produce here are not observed on most of them, our study does not contradict this scenario: the current inclinations of these moons would favor multiple collisions before the final merger, which may produce featureless/random shapes. Moreover, the mass of these objects is large enough that their shapes have evolved closer to hydrostatic equilibrium. Finally, these moons are older and it is difficult to infer if the pyramidal regime would have been the last process to significantly affect their shapes.

Interestingly, Iapetus is the only large moon that displays an oblate shape and possesses an equatorial ridge, although this moon was not proposed to have formed in the pyramidal regime[7]. Its ridge was previously suggested to have formed either by endogenic or exogenic processes[24-29]. To explore if these features can result from the merging of similar-sized bodies, we performed additional collisional simulations with sizes and densities adjusted for the Iapetus case (Methods). We note that tidal forces are negligible at Iapetus' location. Our modeling results suggest that Iapetus' oblate shape as well as the equatorial ridge may be a result of a merger of similar-sized moons, taking place with velocities around $V_{imp}$ = 1.2-1.5 $V_{esc}$ and a close to head-on impact angle (Figure 4). The probability for such an impact is difficult to estimate, as it may not have occurred on the current orbit of the moon. We note, however, that if the peculiar shape of Iapetus is due to the merging of similar-sized bodies, possibly after a series of hit-and-run encounters, its precursors must have been on a similar orbital plane. In that case Iapetus probably formed before gaining its significant inclination with respect to Saturn's equatorial plane, or out of bodies that had another common dynamical origin[30].

# References

1. Thomas, P.C. Sizes, shapes, and derived properties of the saturnian satellites after the Cassini nominal mission. *Icarus* **208**, 395–401 (2010).
2. Charnoz, S., Brahic, A., Thomas, P. C. & Porco, C. C. The Equatorial Ridges of Pan and Atlas: Terminal Accretionary Ornaments? *Science* **318,** 1622-1624 (2007).
3. Porco, C. C., Thomas, P. C., Weiss, J. W. & Richardson, D. C. Saturn's Small Inner Satellites: Clues to Their Origins. *Science* **318,** 1602 (2007).
4. Charnoz, S., Salomon, J. & Crida, A., The recent formation of Saturn's moonlets from viscous spreading of the main rings. *Nature* **465,** 752-754 (2010).
5. Karjalainen, R., Aggregate impacts in Saturn's rings. *Icarus* **189,** 523-537 (2007).
6. Hyodo, R. & Ohtsuki, K., Saturn's F ring and shepherd satellites a natural outcome of satellite system formation. *Nature Geoscience* **8,** 686-689 (2015).
7. Crida, A. & Charnoz, S., Formation of Regular Satellites from Ancient Massive Rings in the Solar System. *Science* **338**, 1196 (2012).
8. Porco, C. C. et al. Cassini Imaging Science: Initial Results on Phoebe and Iapetus. *Science* **307,** 1237-1242 (2005).
9. Thomas, P. C. et al. Shapes of the saturnian icy satellites and their significance. *Icarus* **190,** 573-584 (2007).
10. Dombard A. J., Cheng, A. F., McKinnon, W. B. & Kay, J. P. Delayed formation of the equatorial ridge on Iapetus from a subsatellite created in a giant impact. *JGR* **117,** CiteID E03002 (2012).
11. Goldreich, P. & Tremaine, S. Disk-satellite interactions. *Astrophys. J.* **241**, 425-441 (1980).
12. Jutzi, M. & Asphaug, E. The shape and structure of cometary nuclei as a result of low velocity collisions. *Science* **348**, 1355 (2015).
13. Wisdom, J. The resonance overlap criterion and the onset of stochastic behavior in the restricted three-body problem. *Astron. J.* **85**, 1122-1133 (1980).
14. Deck, K. M., Payne, M. & Holman, M. J. First-order Resonance Overlap and the Stability of Close Two-planet Systems. *Astrophys. J.* **774**, article id. 129 (2013).
15. Poulet, F. & Sicardy, B. Dynamical evolution of the Prometheus-Pandora system. *Mon. Not. R. Astron. Soc.* **322**, 343-355 (2001).
16. Benz, W. & Asphaug, E. Simulations of brittle solids using smooth particle hydrodynamics. *Computer Physics Communications* **87**, 253-265 (1995).
17. Jutzi, M., Benz, W. & Michel, P. Numerical simulations of impacts involving porous bodies. I. implementing sub-resolution porosity in a 3D SPH hydrocode. *Icarus* **198**, 242-255 (2008).
18. Jutzi, M. SPH calculations of asteroid disruptions: the role of pressure dependent failure models. *Planetary and Space Science* **107**, 3 (2015).
19. Hyodo, R. & Ohtsuki, K. Formation of Saturn's F ring by collision between rubble-pile satellites.American Astronomical Society, DPS meeting #46, id.413.11 (2014).
20. Asphaug, E., Agnor, C. B., & Williams, Q. Hit-and-run planetary collisions. *Nature* **439**, 155 (2006).
21. Richardson, D. C. Tree Code Simulations of Planetary Rings. *Mon. Not. R. Astron. Soc.* **269,** 493 (1994)
22. Thomas, P. C. Sizes, shapes, and derived properties of the saturnian satellites after the Cassini nominal mission. *Icarus* **208,** 395-401(2010).
23. Hamilton, D. P., Proctor, A. L., Rauch; K. P. An explanation for the high inclinations of Thebe and Amalthea. *Bulletin of the American Astronomical Society* **33** 1085 (2001).
24. Castillo-Rogez, J. C. et al. Iapetus' geophysics: Rotation rate, shape, and equatorial ridge. *Icarus* **190,** 179-202 (2007).
25. Singer, K. N. & McKinnon, W. B. Tectonics on Iapetus: Despinning, respinning, or something completely different? *Icarus* **216,** 198-211 (2011).

26. Ip, W.-H. On a ring origin of the equatorial ridge of Iapetus. *Geophys. Res. Lett.* **33**(16), L16203 (2006).
27. Levison, H.F., Walsh, K.J., Barr, A.C., & Dones, L. Ridge formation and de-spinning of Iapetus via an impact-generated satellite. *Icarus* **214**, 773–778 (2011).
28. Dombard, A. J., Cheng, A. F., McKinnon, W. B. & Kay, J. P. Delayed formation of the equatorial ridge on Iapetus from a subsatellite created in a giant impact. *Journal of Geophysical Research* **117,** CiteID E03002 (2012).
29. Stickle, A. M. & Roberts, J. H. Building a Ridge That Iapetus Pays For. 48th Lunar and Planetary Science Conference, Texas, Contribution No. 1964, id.1262 (2017).
30. Asphaug, E. & Reufer, A. Late origin of the Saturn system. *Icarus* **223,** 544-565 (2013).

**Acknowledgements** The authors acknowledge support from the Swiss NCCR PlanetS and the Swiss National Science Foundation.

**Author contributions** A.L. performed the dynamical modelling and analyzed the results. M.J. performed the collision modelling and analyzed the results. M.R. contributed initial ideas for the study. All authors contributed to interpretation of the results and the preparation of the manuscript.

**Author Information** Reprints and permissions information is available at www.nature.com/reprints. The authors declare no competing financial interests. Correspondence and requests for materials should be addressed to adrien.leleu@space.unibe.ch.

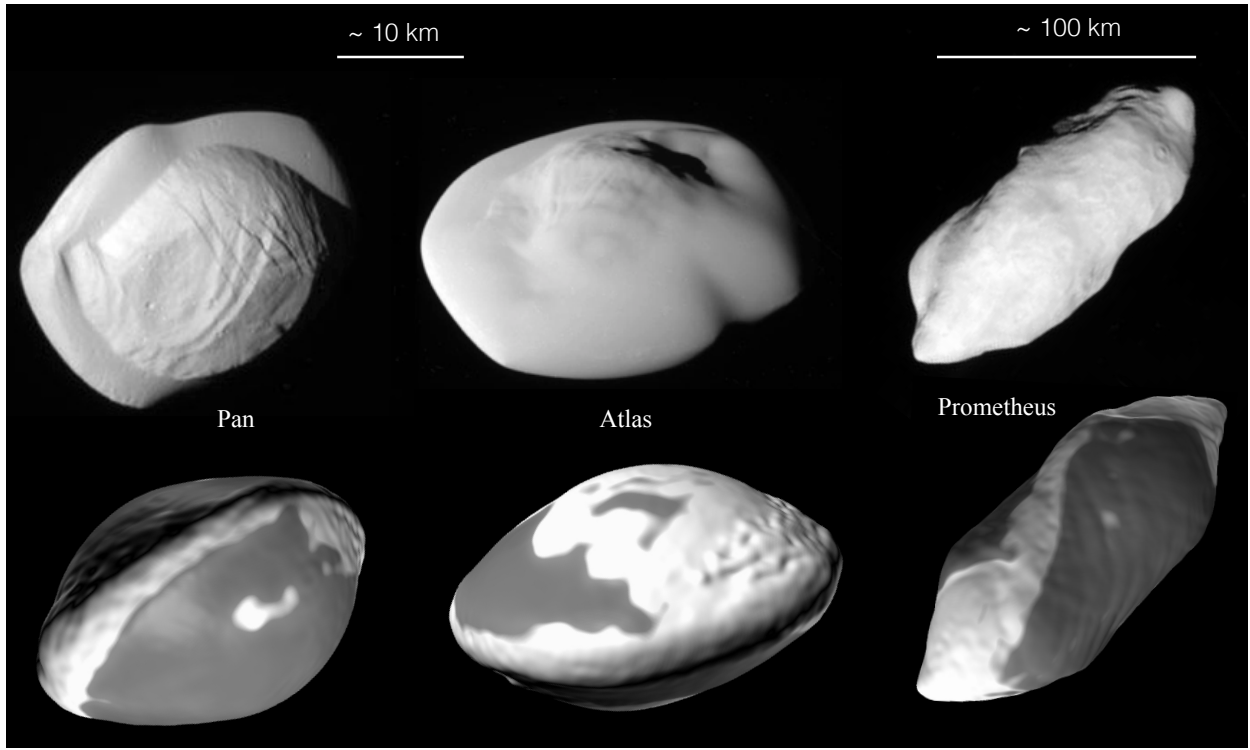

**Figure 1: Shapes of small moons of Saturn compared to model outcomes.** Top: Cassini observations (CREDIT: NASA/JPL-Caltech/Space Science Institute) of Pan and Atlas revealed pronounced equatorial ridges. The elongated structure of Prometheus shows very different characteristics. Bottom: Results of SPH calculations of low-velocity mergers of similar-sized moonlets. Left: $V_{imp} = 2\ V_{esc}$, $\theta = 2°$, $m_1/m_2 = 1$; middle: $V_{imp} = 2.25\ V_{esc}$, $\theta = 0°$, $m_1/m_2 = 1$; right: $V_{imp} = 2\ V_{esc}$, $\theta = 6°$, $m_1/m_2 = 1$. White regions correspond to highly strained / ejected material (total integrated strain >1), fine-grained material, while dark regions indicate less affected, original surface.

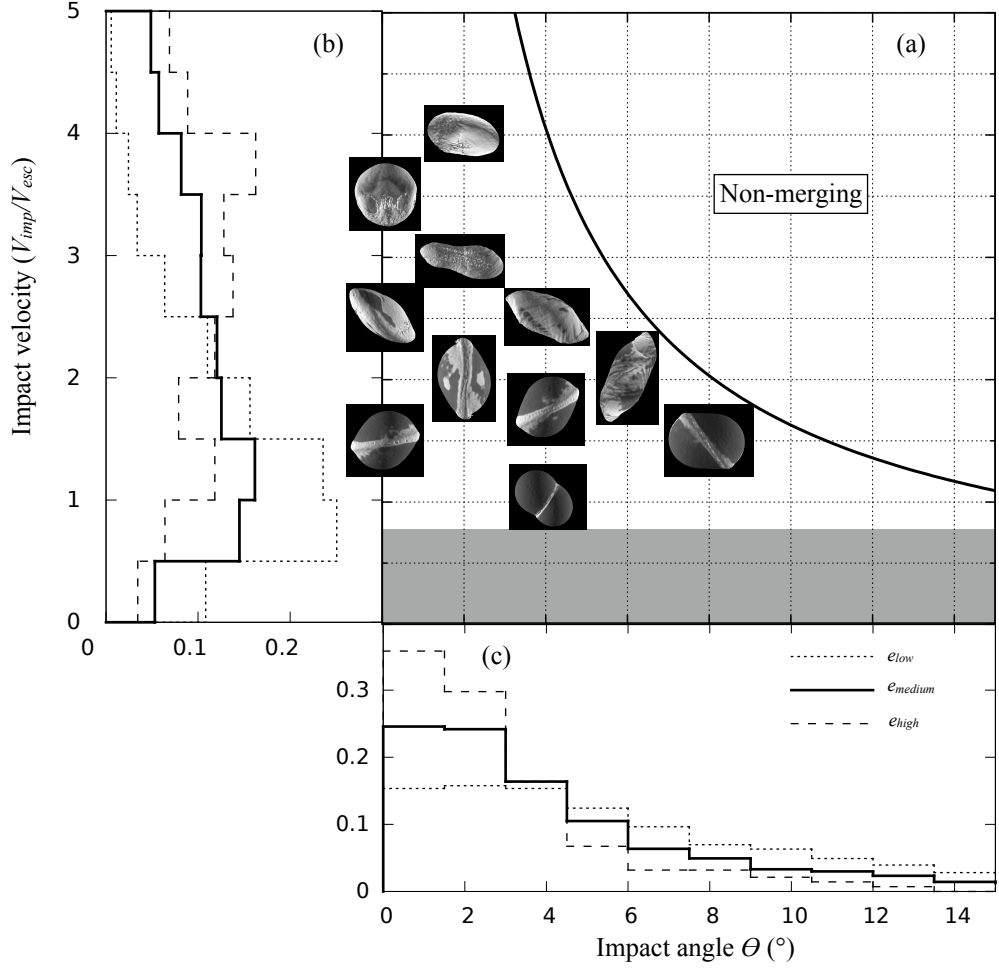

**Figure 2: Impact probability distribution and collision regimes.** The solid line in panel (a) marks the transition between merging / non-merging collisions, defined as $V_{imp}/V_{esc} = 0.4 \sin(45°)/\sin(\theta[°])$, corresponding to a constant initial angular momentum[12]. Examples of shapes resulting from collisional mergers (Figure 3; both mass ratios) are included. At very low impact speeds (<~ 0.75 $V_{esc}$, gray area), distinct bilobate-type shapes are produced. These configurations are unstable and either separate due to tidal and Coriolis forces into two components, or might evolve into elongated bodies. The evolution of $10^4$ pairs of moonlets of mass $m_1/m_2=6/5$ and $m_1+m_2=m_{Atlas}$ have been studied for 3 different distributions of initial eccentricities. Panels (b) and (c) show, respectively, the distributions in $V_{imp}/V_{esc}$ and $\theta$ for the moonlets that merged with $m_1/m_2<7/4$, resulting in shapes comparable to those presented in panel (a). The fractions that reached $m_1/m_2>7/4$ before the end of the simulation are not represented and we assume that the significant disruption/reaccretion they undergo results in random shapes. 30% of the systems with the $e_{medium}$ distribution (equivalent to $10^5$ years spent in the chaotic area, the mean value of the distribution being close to the current eccentricity of Atlas $<e_{medium}> \sim e_{Atlas}$) merged with $m_1/m_2<7/4$, 50% for $e_{low}$ (equivalent to ~$10^4$ years in the chaotic area, $<e_{low}> \sim e_{Atlas}/2$), and 18% for $e_{high}$ (equivalent to ~$10^6$ years, $<e_{high}> \sim 3e_{Atlas}/2$).

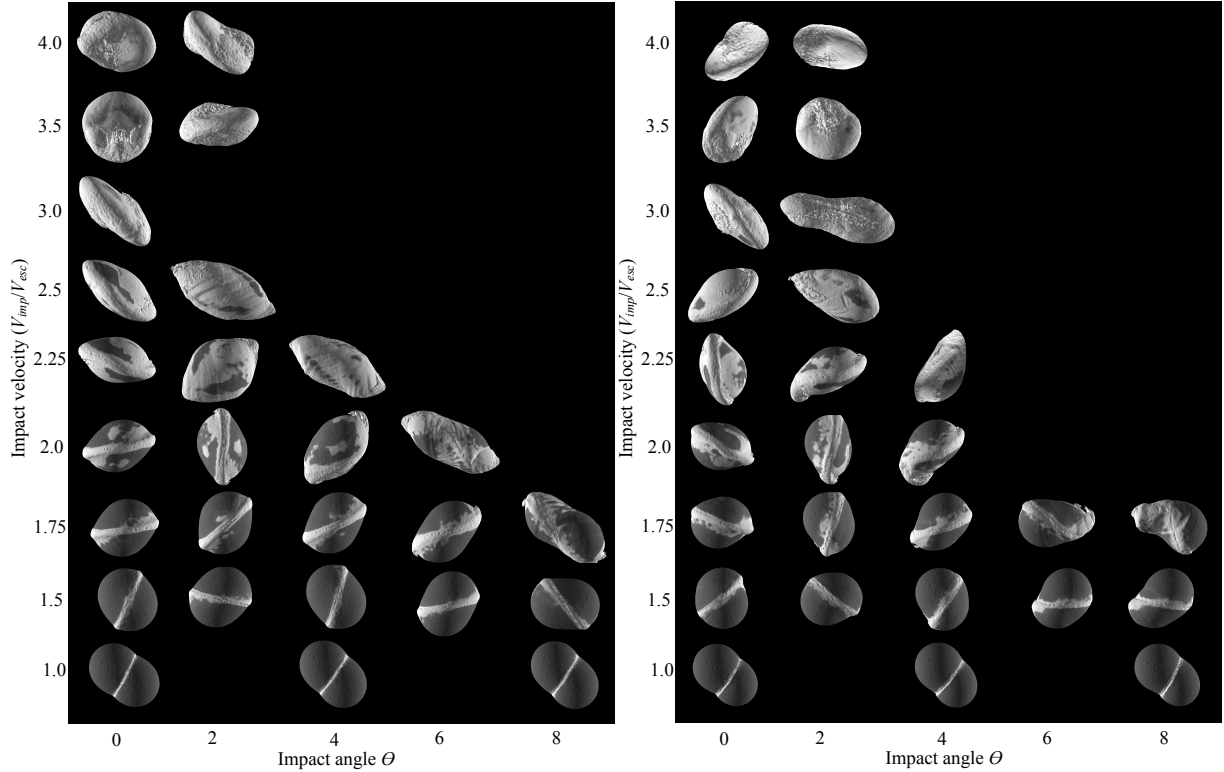

**Figure 3**: **Shapes resulting from merging collisions.** Shown are the results of SPH impact code calculations of low-velocity mergers among similar-sized moonlets, taking place within the Roche distance of Saturn. Collisions of bodies with an initial mass ratio of $m_1/m_2 = 1$ (left) and $m_1/m_2 = 1.5$ (right) are investigated for a range of impact angles $\theta$ and velocities $V_{imp}/V_{esc}$. The calculations start with initially spherical objects modelled as porous aggregates with initial bulk densities of $\rho = 700$ kg/m$^3$ and include self-gravitation as well as tidal forces. The resulting shapes are displayed as iso-density surfaces ($\rho_{iso} = 500$ kg/m$^3$). White regions correspond to highly strained / ejected material (total integrated strain > 1), smoothed material; while dark regions indicate less affected, original surface. The simulations were carried out to $t \sim 17$ h (>= one orbital period). At this time the objects produced in the collisional mergers are not yet aligned with the orbital plane, but will be reoriented later on due to the torques resulting from the tidal forces.

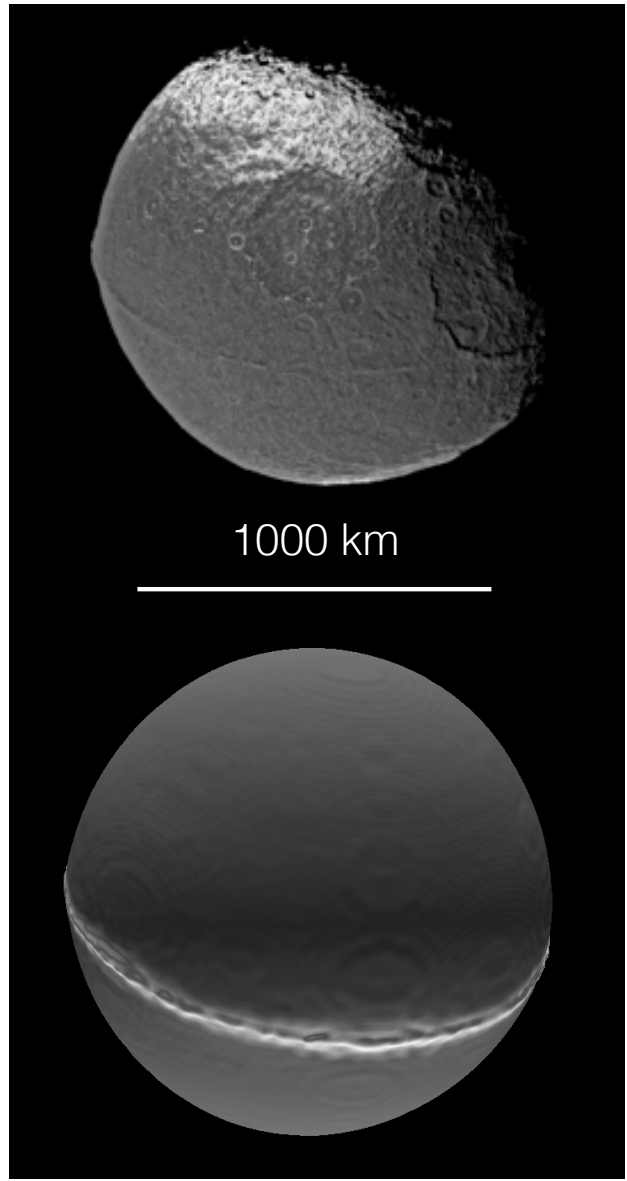

**Figure 4: Ridge formation on Iapetus.** Top: Iapetus as observed by Cassini (Credit: NASA/JPL/Space Science Institute) has an oblate spheroid shape and an equatorial ridge. Bottom: Shown is the result of an SPH calculation of a head-on merger of two equal-sized bodies with half of the mass of Iapetus taking place with a velocity of $V_{imp}/V_{esc}$ = 1.25, leading to a distinct equatorial ridge and an overall shape with an oblateness of the order of a few %, comparable with the observed one (~ 4%).

**Methods**

**Orbital trajectories:**

*Implementation of the forces*

The interaction of two moonlets orbiting Saturn was determined by the integration of the equation of the 3-body problem in astrocentric coordinates, where the potential of the center body was modified to take into account the first harmonic of the oblateness ($J_2$) of Saturn[31]:

$$V_{Sat}(r,\varphi) = -\frac{Gm_{Sat}}{r}\left(1 - \frac{J_2}{2}\left(\frac{R_{Sat}}{r}\right)^2 (3\cos^2\varphi - 1)\right),$$

where $\varphi$ is the colatitude, r is the distance from the moonlet to the center of Saturn, $R_{Sat}$ is the equatorial radius of Saturn and $G$ is the universal constant of gravitation. The integrations were performed using the variable-step integrator DOPRI (Runge-Kutta 8(7)).

*Secular perturbation from the main moons*

The secular perturbations induced by one of the main Saturnian moons on a moonlet near the rings of Saturn can be estimated by the Laplace-Lagrange solution of the secular problem generalized to take into account the oblateness of the central body[31]. Considering the first order in $J_2$, we checked the effect of Mimas (the innermost main moon) and Titan (the largest), on both Atlas (closest to Saturn) and Epimetheus (the outermost considered small moon); in each case, the forced eccentricities due to the secular perturbation are lower than $10^{-5}$. The effects of the main moons can hence be neglected as long as the considered semi-major axes are not too close to the 3:2 and 4:3 mean-motion resonances with Mimas (currently located at ~141500 km and ~153000 km, respectively).

*Eccentricity excitation in the chaotic area*

In the case of the coplanar 3-body problem, for initially circular orbits, the moonlets will be in the chaotic area due to the overlap of first-order mean-motion resonances (MMR) if their semi-major axis $a_j$ satisfy[13,14] ($\mu=(m_1+m_2)/m_{Sat}$):

$$|a_1 - a_2|/a_1 \leq 1.5\, \mu^{\frac{2}{7}},$$

and they are not protected by the 1:1 (co-orbital) MMR. At $a_1=a_{Atlas}$, this yields a half-width of ~150km for $m_1+m_2=m_{Atlas}$, ~385 km for $m_1+m_2=m_{Prometheus}$, and ~860 km for $m_1+m_2=m_{Janus}$. The precession induced by the oblateness of Saturn displaces individual MMRs and might split them into corotation and Lindblad resonances that could enlarge the chaotic area[32].

To estimate the orbital evolution of two moonlets in the chaotic area, we integrate the trajectories of a pair of moonlets for different initial values of $a_1-a_2$. The two moonlets are initially placed on circular orbits such that the semi-major axis of the outer moonlet is the current semi-major axis of Atlas, and the initial mean longitudes were chosen randomly. The maximum time step is set to 1/20 of the orbital period, and the integrator precision to $10^{-14}$. The results are presented in Supplementary Figure 1.

The graphs (a), (b), and (c) represent the evolution of the eccentricity (color code in log scale) of the body $m_1$ as a function of the time and the initial value of $a_1-a_2$, in the case $m_1=m_2$. Three cases are represented: $m_1+m_2=m_{Atlas}$ (a), $m_1+m_2=m_{Prometheus}$ (b), and $m_1+m_2=m_{Janus}$ (c). For these objects on initially circular orbit, the stable 1:1 mean-motion area extend roughly up to $a_1-a_2 \sim 0.7 a_1 \mu^{1/3}$, but decrease for larger eccentricities[33]. The figure (d) represent the mean value of the eccentricity of $m_1$ after a given time spent in the chaotic area for each pair of half-moons. On average, eccentricities comparable with the current eccentricities of these moonlets is reached within $10^5$ years.

The graph (e) represents the same quantity as in the graph (d), but with $1/t$ and $<e_1>$ now normalized by $\mu^{1/3}$. We found that the average eccentricities are excited proportional to $\mu^{1/3}$ after a time $\mu^{1/3}t$ spent in the chaotic area, a scaling that is present in the Hamiltonian of individual resonances at low eccentricities[34].

*Estimation of the ratio $V_{imp}/V_e$*

The mutual escape velocity of two moonlets of equal mass $m_1=m_2=\mu m_{Sat}/2$ is given by:

$$V_{esc} = \sqrt{\frac{2G(m_1+m_2)}{R_1+R_2}} = \sqrt{2G\left(\frac{4\pi\rho}{3}\right)^{1/3}} m_{Sat}^{1/3} \left(\frac{\mu}{2}\right)^{1/3},$$

where $R_1$ and $R_2$ are the radius of $m_1$ and $m_2$ in the case of spherical bodies and $\rho$ is the mean density of the moonlets. At the considered densities ($\rho \sim 500 - 700$ kg/m$^3$) the Hill radius of the moonlets is comparable to their physical radius; the impact velocity $V_{imp}$ is hence comparable to the relative velocity. At a given eccentricity, the impact velocity can be estimated by:

$$V_{imp} \approx V_{relative} \approx nae = \sqrt{Gm_{Sat}/a}\,e.$$

We hence obtain:

$$\frac{V_{imp}}{V_{esc}} \approx m_{Sat}^{\frac{1}{6}} \left(\frac{4\pi\rho}{3}\right)^{-\frac{1}{6}} \frac{1}{2^{\frac{1}{6}}\sqrt{a}} \frac{e}{\mu^{1/3}}.$$

Taking the semi-major axis of the considered small moonlets outside the main ring $a \sim 137500$ km and $\rho = 700$ kg/m³, we obtain $V_{imp}/V_{esc} \sim 0.58 e/\mu^{1/3}$.

**Collision modeling**

To model the collisions we use a smooth particle hydrodynamics (SPH) impact code[35,17,18]. This code includes self-gravity and is especially suited to modeling collisions between rocky/granular bodies. We assume that the moonlets are spherical, cohesionless porous low-density aggregates. In the model, a pressure dependent shear strength (friction) is included by using a standard Drucker-Prager yield criterion[18]. Granular flow problems (of cohesionless material) are well reproduced using this method[18,36]. A coefficient of friction $\mu = 0.8^{37}$ is used. Porosity is modeled using a P-alpha model with crush-curve parameters corresponding to pumice[38]. We use the Tillotson EOS with a reduced bulk modulus of $A = 2.7 \times 10^8$ Pa (leading term in the EOS) to take into account the smaller elastic wave speeds in porous materials compared to solid rock. We note that for the impact velocities considered here (up to a few 10 m/s), no significant heating occurs and the bulk modulus together with the crush-curve parameters solely define the material response in compression.

$5 \times 10^5 - 10^6$ SPH particles are used in the simulation of the merging collisions. In order to explore the parameter space of non-merging collisions, low resolution runs with $\sim 5 \times 10^4$ particles are performed.

To include tidal effects, we use the linearized equation of motion in a rotating Cartesian coordinate system[39] orbiting around Saturn. The colliding small moonlets are assumed to have a tidally locked rotation state, i.e., they have zero spin in the rotating coordinate system.

We use an $x$-axis pointing away from Saturn, the y-axis tangential to the direction of motion, and the $z$-axis perpendicular to the orbital plane. This yields the following forces per unit mass, which are added as external accelerations in each time-step to the SPH particles:

$dx^2/dt^2 = 3\ \Omega^2 x + 2\ \Omega\ dy/dt$

$dy^2/dt^2 = -\ 2\ \Omega\ dx/dt$

$dz^2/dt^2 = -\ \Omega^2\ z$

where $\Omega^2 = G\ M_{Saturn} / a^3$, G is the gravitational constant, and $a$ is the distance to Saturn. In our nominal case $a$ is set to $1.33 \times 10^8$ m and the impact geometry is oriented so that the bodies are moving along the x-axis prior to the collision. The effects of the strength of the tidal force as well as the orientation are shown in Supplementary Figure 2.

Previous studies of small body collisions suggest that structures resulting from large-scale collisions may have an increased porosity[38,39]. We therefore start with bodies with initial bulk densities ($\rho = 700$ kg/m$^3$) which are slightly larger than those observed for the small moons ($\rho \sim 500$ kg/m$^3$). It has been suggested that the small moons formed from accretion of ring debris around a denser core of $\rho \sim 900$ kg/m$^3$, but the shapes of the moons are also consistent with a homogenous structure[3]. Here, we assume an intermediate density (and initially homogenous structure) for the colliding moonlets. The structures resulting from merging collisions have non-homogenous density distributions and average densities lower than the initial ones, as discussed below.

We first investigate the possible evolution of the spherical shapes of the individual bodies under the influence of self-gravitation and tidal forces. We find that these bodies, supported by the pressure-dependent shear strength (friction), keep their spherical shapes, and we therefore use initially spherical bodies in our simulations.

After non-merging hit-and-run collisions, the bodies may not be perfectly spherical any more, which could affect the outcome of the subsequent merger. Due to computational limitations, it is not possible to systematically track and compute all the shapes resulting from hit-and-run events. We compute for one characteristic case (with $m_1/m_2 = 1.125$, $V_{imp}/V_{esc} = 2$, $\theta = 45°$) the post hit-and-run shape of the largest remnant. This shape is then used to study the outcome of a merger of two irregularly shaped bodies (using $m_1/m_2 = 1$, $V_{imp}/V_{esc} = 2$, $\theta = 90°$). We find that the main features (large-scale ridges) are similar as in the case of a merger of spherical bodies with otherwise the same initial conditions (Supplementary Figure 3).

**Post-impact analysis**

*Properties of merged bodies*

We compute the total strain experienced by the material during the final collision by integrating the second invariant of the strain-rate tensor for each SPH particle (over the entire simulation). We use this measure to distinguish between highly sheared (integrated strain > 1), possibly fined-grained material, and material which was less affected (integrated strain < 1), and kept its original properties. Material that was ejected and reaccumulated on the final merged structure has undergone high strains (>> 1). We note that the potential (small) deformations experienced during the previous hit-and-run events are not considered in these calculations.

As found in previous studies[12], the low velocity collisions considered here do not lead to any significant compaction. However, several mechanisms, such as shear dilatation or the ejection of material followed by reaccumulation, can lead to a final bulk density which can be significantly lower than that of the colliding bodies[40,41].

To estimate the final density of the merged structure, we assume that the highly strained material (>1) experienced an increase of porosity either due to shear dilatation[44] or the process of reaccumulation which introduces macroporosity[40,41]. The corresponding material is assigned an increase of distention by 1/(1-0.4)=~1.7, following ref (40). The resulting average densities depend on the impact parameters (the density generally decreases with increasing impact velocity) and the resulting structure. For initial densities of $\rho$ = 700 kg/m$^3$, the structures formed with $V_{imp}/V_{esc}$ >~ 1.75 (Figure 3) are in the range of 400-600 kg/m$^3$.

We note that this analysis is performed as a post-processing step; the bodies shown in Figures 1-4 are based on the original densities. As a result, the material shown in white would have a larger volume and therefore the ridges would be slightly more pronounced.

*Change of mass ratio in collisions*

In the case of the non-merging collisions, the post-impact mass ratio *mr'* depends on the initial mass ratio *mr* as well as the impact parameters. We perform a set of simulations using different initial mass ratios to determine the change of the mass ratio in non-merging collisions as a function of impact angle and velocity. This allows for an empirical determination of *mr'*=*f*(*mr*) in different areas of the parameter space with varying degree of interaction.

The post-impact mass ratio is estimated by computing the masses of the largest two remnants after separation, considering SPH particles which are part of the individual bodies. For this, all accreted SPH particles with $\rho / \rho_0$ > 0.7 (where $\rho_0$ is the initial bulk density) are considered in the calculation of the mass; the lower density ejecta are not counted as part of the body.

While the transition from merging to non-merging collisions is relatively well defined by the angular momentum given by the collision geometry and velocity, the mass ratios resulting from the non-merging collisions have a more complex dependence on the impact parameters. In addition to the angular momentum, the strong tidal and Coriolis forces lead to a highly non-linear behavior. Collisions close to the merging / non-merging boundary typically lead to an initial merging of the two bodies, resulting in an unstable configuration which leads to a subsequent splitting into two components. On the other hand, collisions taking place far from the transition line show the characteristics of a typical hit-and-run event, with not much interaction between the two bodies.

The relative change of the mass ratio in the non-merging collisions is shown in Supplementary Figure 4 as a function of impact angle and normalised velocity. We can roughly define three zones to characterize the relative change of the mass ratio in a non-merging collision, corresponding to different degrees of change.

For each zone, we compute the average relative change of the mass ratio. Supplementary Figure 5 shows the dependence of the relative change of mass ratio as a function of the initial mass ratio for the three different zones. We find that the linear function $(mr'-mr)/mr = ar \, (mr-1)$ reproduces well the dependence on the initial mass ratio in the considered range. Here, $mr = m_1/m_2$ is the mass ratio before the collision and $mr' = m_1'/m_2'$ is the mass ratio after the collision. The best fit values are $ar = 0.11$ (zone 1); $ar = 0.65$ (zone 2); and $ar = 2.81$ (zone 3) . Using the linear function as defined above for the different zones, we can compute the post-collision mass ratio as follows:

$mr' = mr \, (ar \, (mr - 1) + 1)$

**Merging probability**

*Hit-and-runs*

The collisions occurring in the hit-and-run regime are treated as inelastic collision with dissipation[21] with a damping factor of 0.8 in the normal direction and 1 in the tangential direction, typical for the collision of rubble piles of the considered density[21,42]. The new mass ratio between the moonlets is computed using the relations given in the previous section.

*Initial conditions*

We explore the effect of a different distribution of eccentricity that accounts for different durations spent in the chaotic area. For the cases displayed in Figure 2, $10^4$ initial conditions were randomly chosen as following: the initial eccentricities $e_1$ and $e_2$ follow the distributions $e_1(t=10^k \text{ yr})$ and $e_2(t=10^k \text{ yr})$ that represent the eccentricity distributions of two bodies on initially circular orbits that remained for $t=10^k$ years in the chaotic area. It is the mean values of these distributions that are represented in Supplementary Figure 1 (d). The semi-major axes are then randomly chosen in order for the moonlets to be on intersecting orbits: $a_1=a_{Atlas}$, and $a_2$ in $[a_1+R_{Hill}+R_{Impact}, a_1(1+e_1+e_2)+R_{Hill}+R_{Impact}]$. Finally, the mean longitudes and longitude of pericenter are randomly chosen in $[0,2\pi)$.

*Comparison of the Atlas and Prometheus cases*

We verify that the dynamical study is scale-free for a given mean-density and semi-major axis by comparing the outcomes of the cases $m_1+m_2=m_{Atlas}$ and $m_1+m_2=m_{Prometheus}$. The initial distribution of eccentricity is taken at $t=10^5$ years in the case of Atlas (the nominal case described in the paper) and the corresponding $t = (m_{Atlas}/m_{Prometheus})^{1/3} 10^5$ years in the case of Prometheus. $2 \times 10^3$ initial conditions are chosen as described in the previous section for the Prometheus case, to compare with the $10^4$ initial conditions for

Atlas. $R_{impact}$ is set to 66 km for Prometheus. With an initial mass ratio $m_1/m_2=6/5$, 29% of the runs ended up with a mass ratio bellow 7/4, against 30% for Atlas, with roughly identical distributions in $V_{imp}/V_e$ and $\theta$ (see Supplementary Figure 6).

*Effect of inclination*

For a random impact in 3D, the probability function $dP$ for the impact angle $\theta$ is given by[43] $dP=sin(2\theta) d\theta$, giving a maximum probability of impact at $\theta=45°$. Let us adapt the reasoning to the 2D case. Consider a meteoroid approaching a massless sphere of radius $r$. If both objects are coplanar, within a segment of length $2r$, all points of intersection of the path of the meteoroid with a plane perpendicular to its path are equally probable, and the total probability $P$ is given by:

$P = 2r = 1$.

The differential probability with which the meteoroid will pass through a point at a distance $x$ from the centre of this segment, where $0 \leq x \leq r$, is:

$dP = 2dx$,

and we have:

$x = r \sin(\theta)$.

As $dx = r \cos(\theta) d\theta$, we obtain $dP=cos(\theta) d\theta$, yielding a maximum probability for head-on impact $\theta=0°$.

In terms of impact angle probability, the system can be considered fully 2-dimensional as long as the vertical excursion of the bodies is significantly smaller than their physical size. In the case of Atlas, the current inclination[45] $i_{Atlas}=0.003°$ leads to a vertical excursion of ~7 km from the midplane, which is comparable to the size of the body. This results in a slight shift of the maximum probability of $\theta$ from zero to a few degrees. The Supplementary Figure 7 represents the distributions of $V_{imp}/V_e$ and $\theta$ in the case where the moonlets had inclinations comparable with the current inclination of Atlas prior to collision. The maximum of the distribution in θ is shifted to a few degrees, while $V_{imp}/V_e$ is shifted toward smaller values.

**Iapetus modeling**

We investigate potential ridge-forming collisions at large, Iapetus-sized, scales using a relatively narrow range of mass ratios from 1:1 to 5:4, impact angles ranging from 0 to 10° and velocities $V_{imp}/V_{esc}$ from 1 to 1.5 (Supplementary Figures 8 and 9). We use homogenous nonporous bodies with densities of 1100 kg/m$^3$ and masses of 0.9 10$^{21}$ kg, roughly half that of Iapetus. Although the spatial resolution (~13 km) in our numerical calculation does not allow for a comparison of the small-scale features of the ridge produced with the observed ones[10], our analysis suggests that the current shape as well as the equatorial ridge may be a result of a merger of similar sized moons, taking place with velocities around $V_{imp}$ = 1.2-1.5 $V_{esc}$ and a close to head-on impact angle.

The data that support the plots within this paper and other findings of this study are available from the corresponding author upon reasonable request.

**Additional references**

31. Murray, C. D. & Dermott, S. F. *SOLAR SYSTEM DYNAMICS.* Cambridge, UK: Cambridge University Press (1999).
32. El Moutamid, M., Sicardy, B. & Renner, S. Coupling between corotation and Lindblad resonances in the presence of secular precession rates. *Celestial Mechanics and Dynamical Astronomy* **118,** 235-252 (2014)
33. Leleu, A., Robutel, P., & Correia, A.C.M. On the coplanar eccentric non-restricted co-orbital dynamics. *Celestial Mechanics and Dynamical Astronomy* **130**, 3 (2018).
34. Petit, Antoine C., Laskar, J. & Boué, G. AMD-stability in presence of first order Mean Motion Resonances. *Astron. Astroph.* **607**, A35 (2017).
35. Benz, W. & Asphaug, E. Simulations of brittle solids using smooth particle hydrodynamics. *Computer Physics Communications* **87**, 253-265 (1995).
36. Jutzi, M., Holsapple, K. A., Wuenneman, K., and Michel, P. Modeling asteroid collisions and impact processes. In *Asteroids IV* (P. Michel, F. DeMeo, and W. F. Bottke, eds.), Univ. of Arizona, Tucson. (2015)
37. Collins, G.S., Melosh, H.J., Ivanov, B.A. Modeling damage and deformation in impact simulations. Meteorit. Planet. Sci. **39**, 217–231 (2004).
38. Jutzi, M., P. Michel, K. Hiraoka, A. M. Nakamura, and W. Benz. Numerical simulations of impacts involving porous bodies: II. comparison with laboratory experiments. *Icarus* **201**, 802 (2009).
39. Wisdom, J. & Tremaine, S. Local simulations of planetary rings. *Astron. J.* **95,** 925-940 (1988)
40. Jutzi, M., W. Benz, A. Toliou, A. Morbidelli, R. Brasser. How primordial is the structure of comet 67P/C-G? *Astron. Astroph.* **597**, A61 (2017)
41. Jutzi, M., W. Benz. Formation of bi-lobed shapes by sub-catastrophic collisions**.** *Astron. Astroph.* **597**, A62 (2017).
42. Beauge, C.; Aarseth, S. J. N-body simulations of planetary formation. *Mon. Not. R. Astron. Soc.* **245** 30-39 (1990).
43. Shoemaker, E. M. Interpretation of lunar craters. *PHYSICS and ASTRONOMY of the MOON* Academic press, New York (1962).
44. Collins, G.S. Numerical simulations of impact crater formation with dilatancy. *J. Geophys. Res. Planets* **119**, 2600–2619 (2014).
45. Spitale, J. N.; Jacobson, R. A.; Porco, C. C.; Owen, & W. M., Jr. The Orbits of Saturn's Small Satellites Derived from Combined Historic and Cassini Imaging Observations. *Astron. J.* **132**, 2, 692-710 (2006).

**Supplementary Information**

Title: The peculiar shapes of Saturn's small inner moons as evidence of mergers of similar-sized moonlets.

Autors: A. Leleu, M. Jutzi, and M. Rubin.

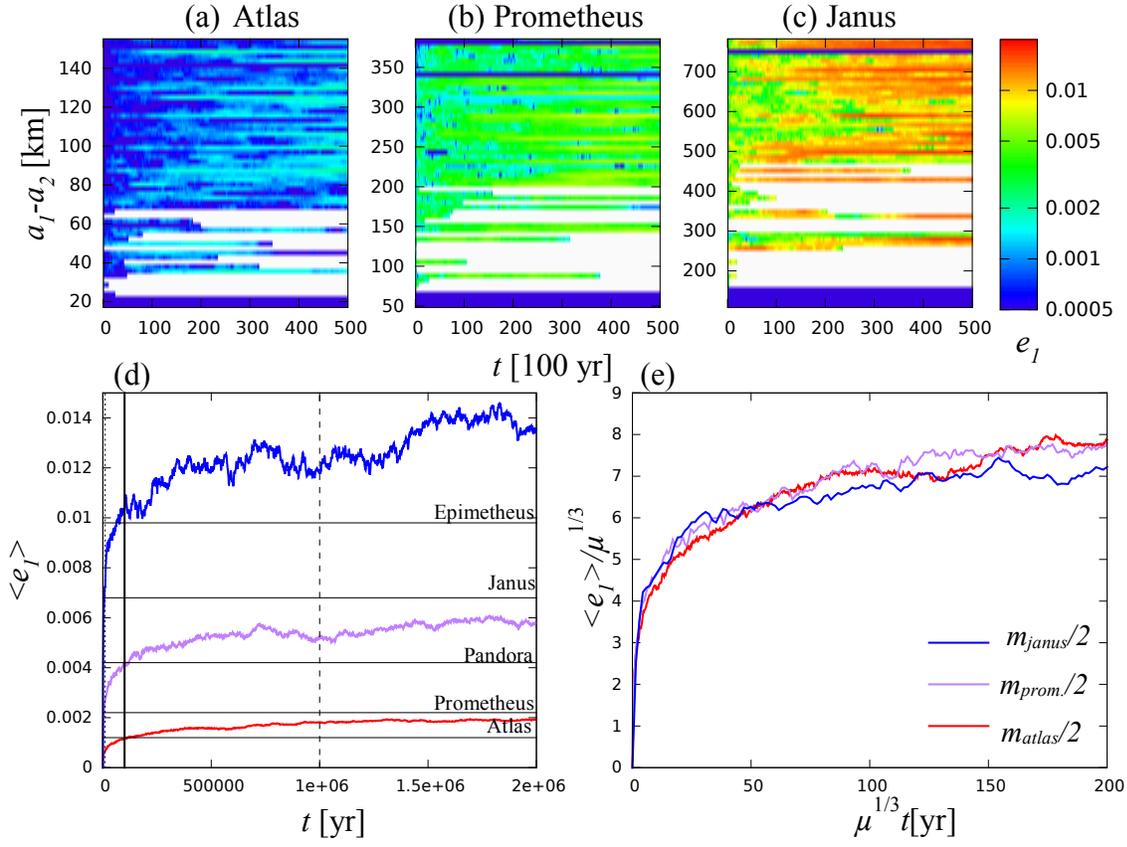

**Supplementary Figure 1: Evolution of the eccentricity of equal-mass moonlets in the chaotic area.**
The cases $m_1=m_2=m_{Atlas}/2$ (a, red in d and e), $m_1=m_2=m_{Prometheus}/2$ (b, purple in d and e), and $m_1=m_2=m_{Janus}/2$ (c, blue in d and e). The graphs (a), (b), and (c) represent the evolution of the eccentricity of $m_1$ over time (x axis) for different initial values of $a_1-a_2$ (y axis). The value of $e_1$ is given by the color code (log scale), and white pixels represent collisions. For $|a_1-a_2|$ below $\sim 0.7\mu^{1/3}$, the moonlets are in a stable co-orbital configuration (as it is currently the case for Janus and Epimetheus) which doesn't lead to a significant increase in eccentricities. Graph (d) represents the evolution of the mean value of $e_1$ over 200 different trajectories spanning the chaotic domain. The horizontal black lines indicate the current eccentricities of the small moons. The same information is displayed in graph (e), where $<e>$ and $1/t$ were normalized by $\mu^{1/3}$.

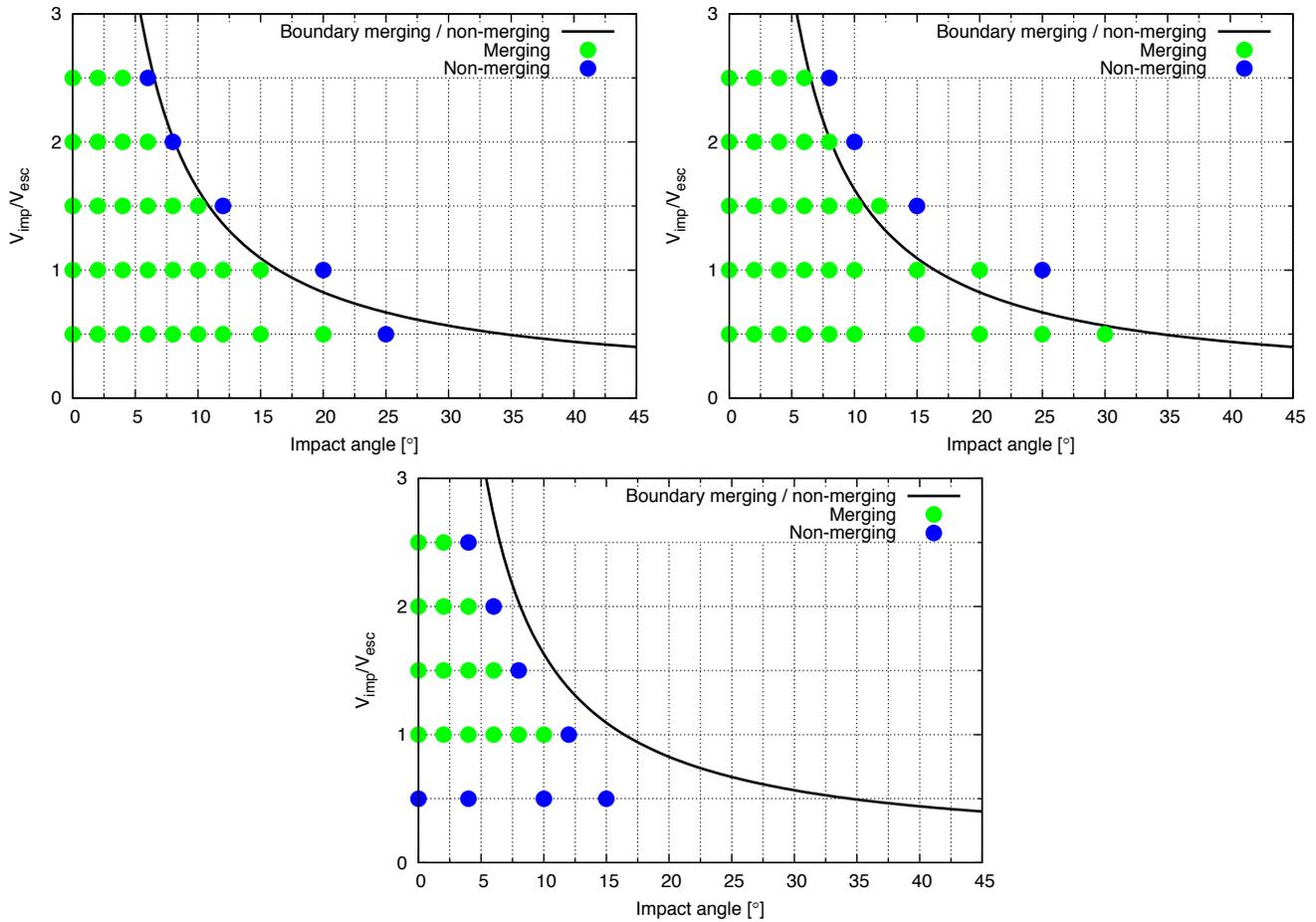

**Supplementary Figure 2: Boundary between merging / non-merging collisions.** Top: Outcome for different tidal strengths. Left: nominal case ($a = 1.33 \times 10^{10}$ m). Right: $a = 1.4 \times 10^{10}$ m (smaller tidal force), leading to an increased merging efficiency. Bottom: Outcome for relative velocities tangent to orbital velocities (using $a = 1.33 \times 10^{10}$ m), leading to a smaller merging efficiency. The nominal case lies in between. The solid line marks the transition between merging / non-merging collisions used in the study.

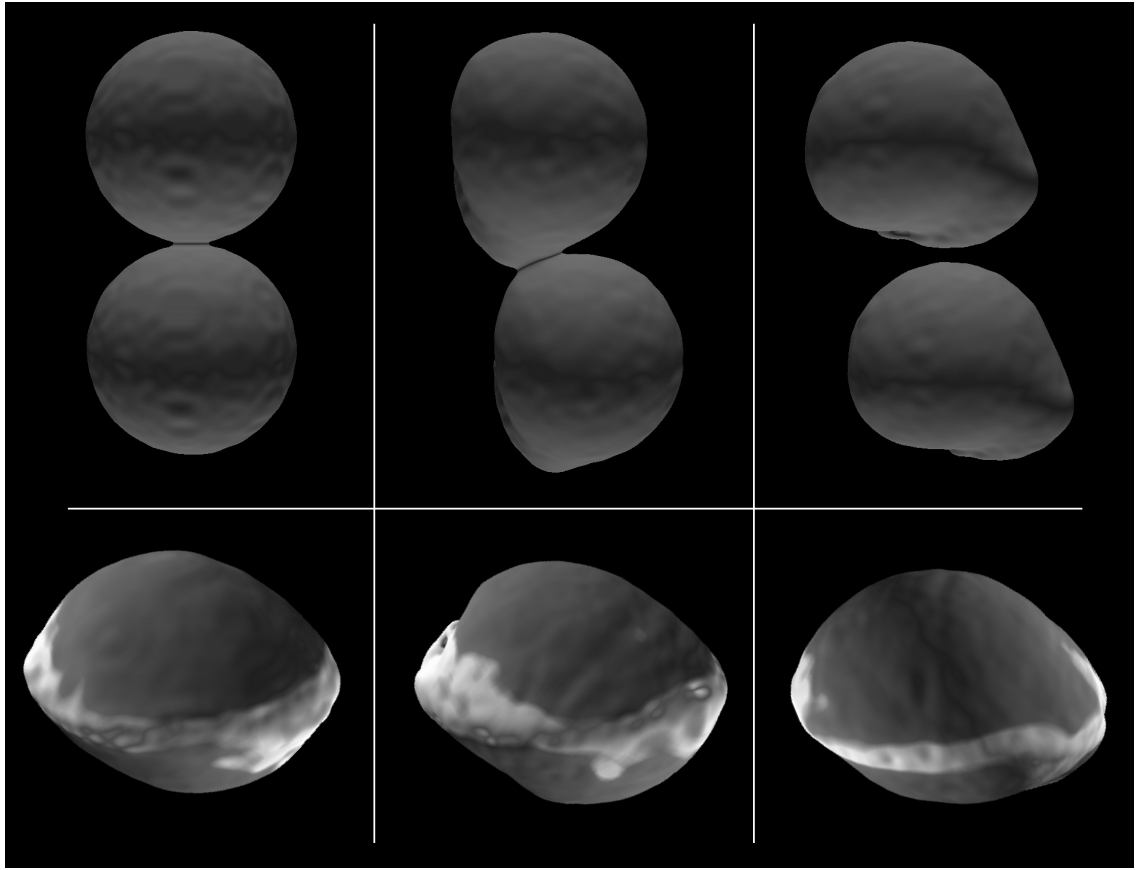

**Supplementary Figure 3: Comparison of collision outcomes using different initial shapes.** Left: spherical bodies. Middle and right: the outcome (largest remnant) of a hit-and-run collision (with $m_1/m_2 = 1.125$, $V_{imp}/V_{esc} = 2$, $\theta = 45°$) is used as the initial shape; two different pre-impact orientations are considered as shown in the top row. The outcome for each case is shown in the bottom row.

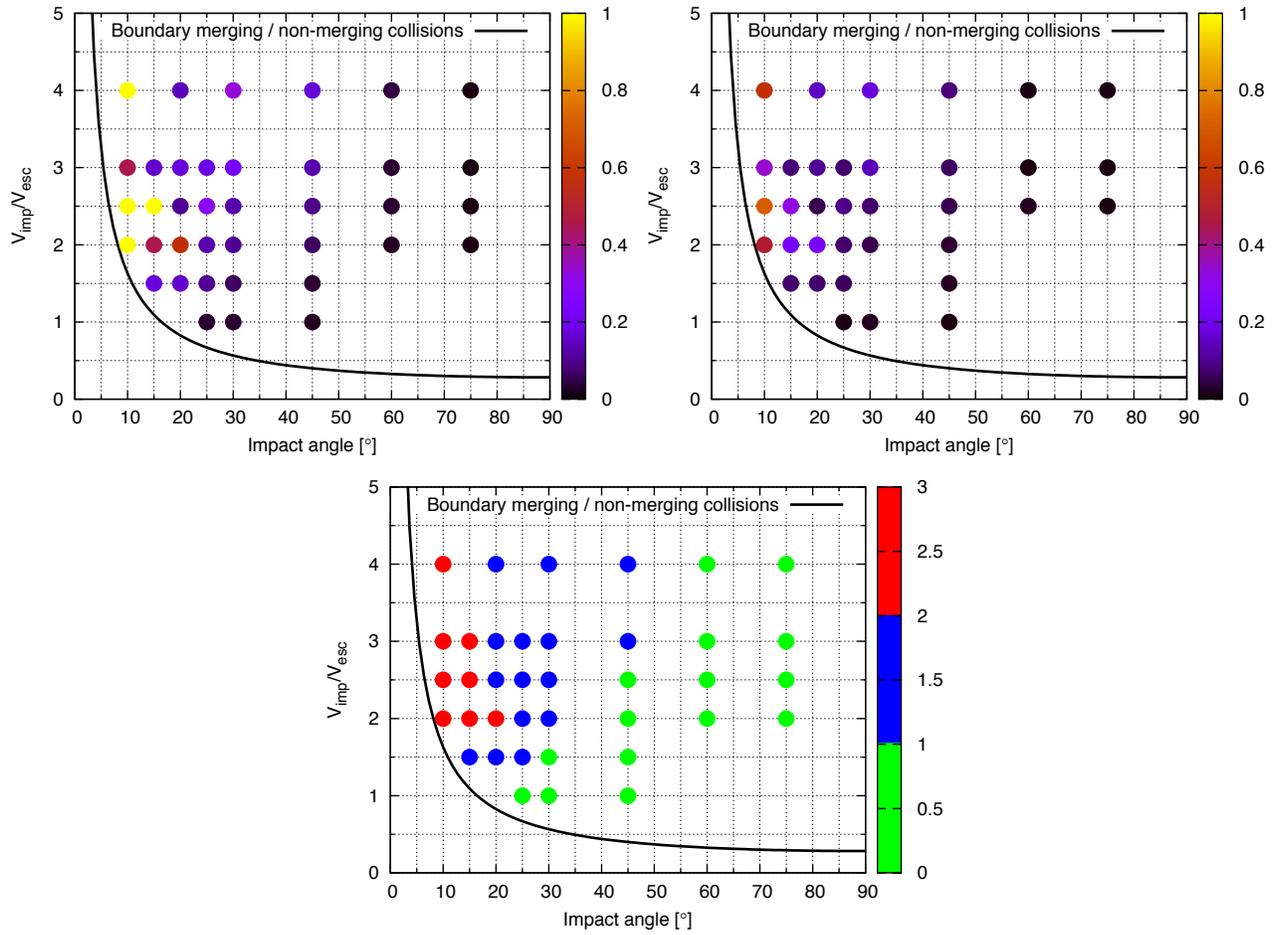

**Supplementary Figure 4: Collision outcome maps.** Shown is the change of the mass ratio as a function of impact angle and velocity. Top left: mass ratio 4:3; top right: 9:8. The average values for the three different zones (green, blue and red) indicated in the plot at the bottom are: 0.026 (green), 0.162 (blue), and 0.725 (red) for the mass ratio 4:3; and 0.015 (green), 0.081 (blue), and 0.308 (red) for the mass ratio 9:8, respectively.

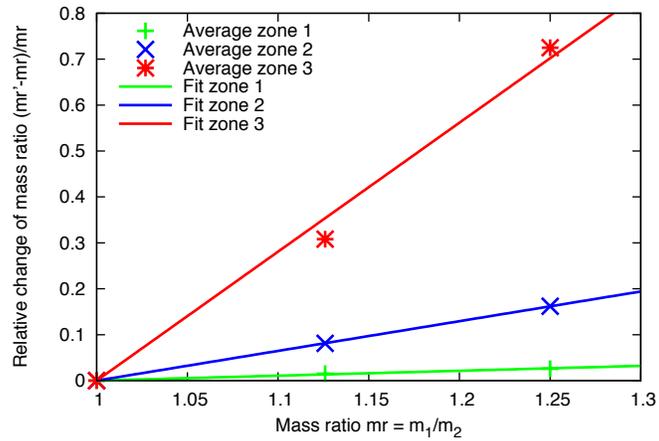

**Supplementary Figure 5: Average relative change of mass ratio in collisions as a function of the mass ratio of the colliding bodies.** The function $(mr'-mr)/mr = ar\,(mr-1)$ is fitted for each zone to determine the slope $ar$, where $mr = m_1/m_2$ is the mass ratio before the collision and $mr' = m_1'/m_2'$ is the mass ratio after the collision. The best fit values are $ar = 0.11$ (zone 1), $ar = 0.65$ (zone 2), and $ar = 2.81$ (zone 3).

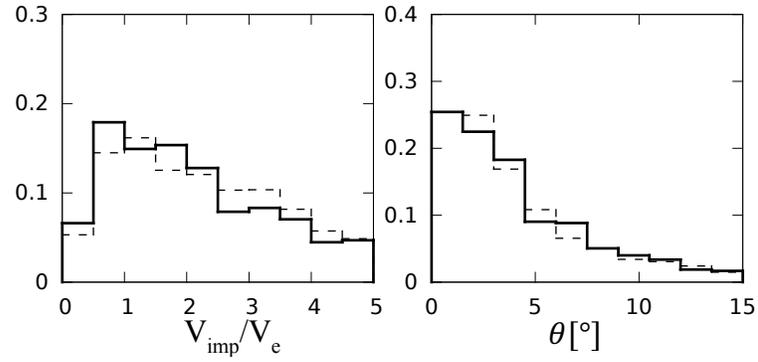

**Supplementary Figure 6**: **comparison of the Atlas and Prometheus cases.** Comparison of the distributions of $V_{imp}/V_{esc}$ and $\theta$ for the final merging of $m_1+m_2=m_{Prometheus}$ (solid lines) and $m_1+m_2=m_{Atlas}$ (dashed lines). Taking $m_1/m_2=6/5$ as initial condition, only the systems retaining $m_1/m_2<7/4$ are displayed here. The initial eccentricity distributions were taken for an equivalent duration of $t/\mu^{1/3}$ spent in the chaotic area.

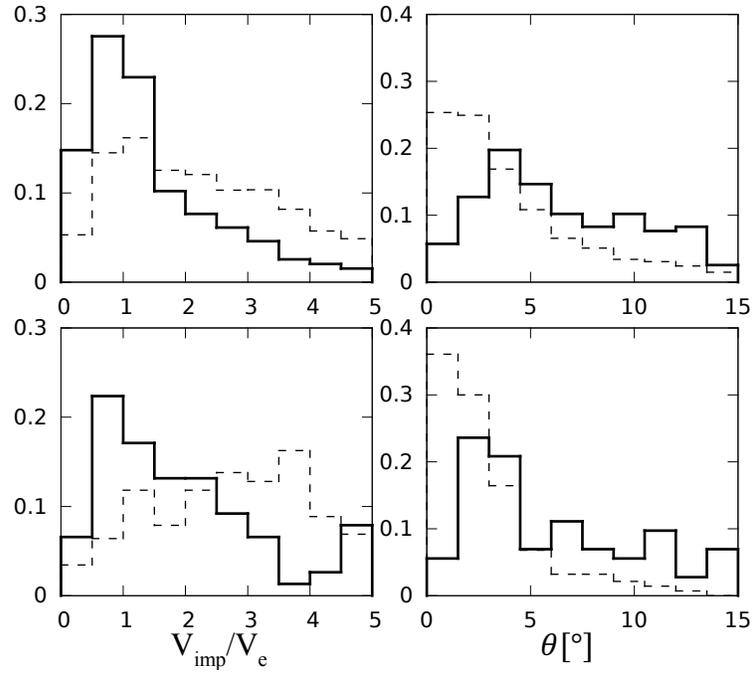

**Supplementary Figure 7: Effect of the inclination**. Comparison of the distributions of $V_{imp}/V_{esc}$ and $\theta$ for the final merging of $m_1+m_2=m_{Atlas}$ and $m_1/m_2=6/5$, with the initial eccentricity distributions $e(t=10^5$ yr$)$ (top) and $e(t=10^6$ yr$)$ (bottom), for the 2D case (dashed lines) and the 3D case (solid lines). Only the systems merging with $m_1/m_2<7/4$ are displayed here. In the 3D case the initial inclinations of $m_1$ and $m_2$ were taken randomly between zero and the current eccentricity of Atlas.

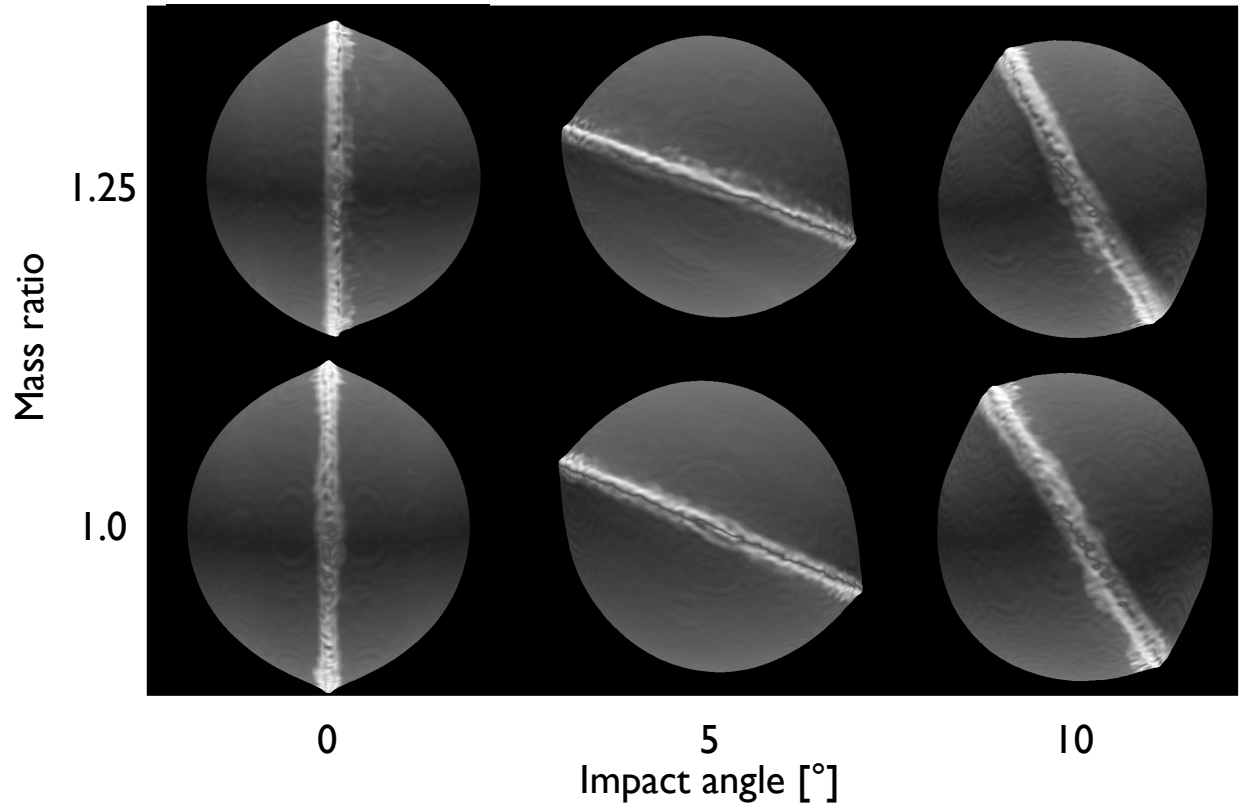

**Supplementary Figure 8:** Merging collisions at large (Iapetus-size) scales, investigating a range of impact angles and mass ratios. The impact velocity is $V_{imp}/V_{esc}$ = 1.375. We note that long-term relaxation (not included in the computation) will lead to shapes closer to hydrostatic equilibrium, which are more akin to that of Iapetus.

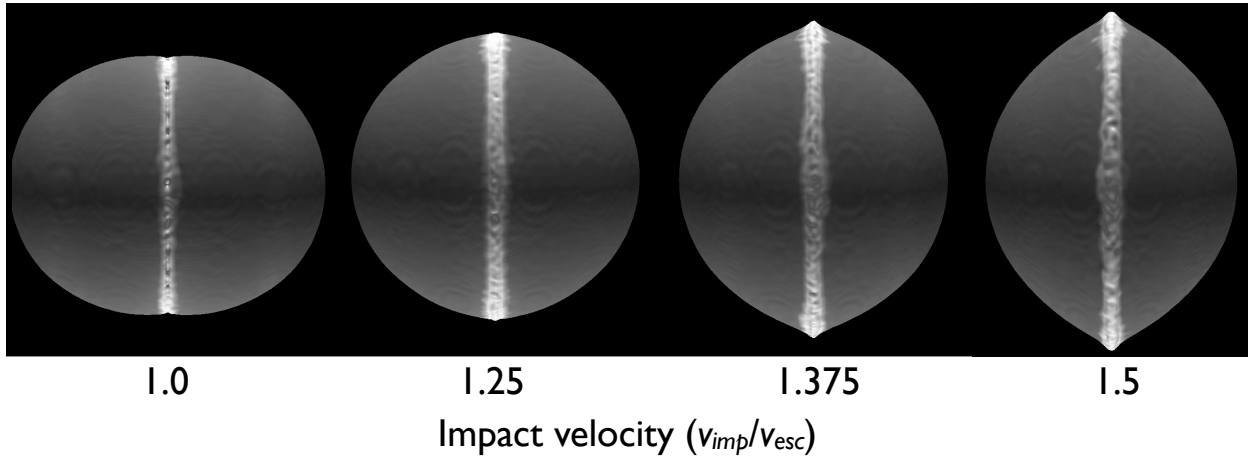

Impact velocity ($v_{imp}/v_{esc}$)

**Supplementary Figure 9:** Same as Supplementary Figure 8**,** but for a fixed mass ratio (1:1), impact angle 90° and varying impact velocities. We note that long-term relaxation (not included in the computation) will lead to shapes closer to the hydrostatic equilibrium, which are more akin to that of Iapetus.